\documentclass[USenglish,twocolumn]{article}
\usepackage[utf8]{inputenc}
\usepackage[big]{dgruyter}

\begin{document}
  \articletype{Research Article{\hfill}Open Access}
  \author*[1]{Anisa Bajkova}
  \author[2]{Vadim Bobylev}
  \affil[1]{Pulkovo Observatory, E-mail: bajkova@gao.spb.ru}
  \affil[2]{Pulkovo Observatory, E-mail: vbobylev@gao.spb.ru}
  \title{\huge Refinement of Parameters of Six Selected Galactic Potential Models}
  \runningtitle{Refinement of Parameters of Six Selected Galactic Potential Models}

  \begin{abstract}
{This paper is devoted to the refinement of the parameters of the six three-component (bulge, disk, halo) axisymmetric Galactic gravitational potential models on the basis of modern data on circular velocities of Galactic objects located at distances up to 200 kpc from the Galactic center. In all models the bulge and disk are described by the Miyamoto--Nagai expressions. To describe the halo, the models of Allen-Santill\'an (I), Wilkinson-Evans (II), Navarro-Frenk-White (III),  Binney (IV), Plummer (V), and Hernquist (VI) are used. The sought-for parameters of potential models are determined by fitting the
model rotation curves to the measured velocities, taking into account restrictions on the local dynamical matter density $\rho_\odot=0.1 M_\odot$~pc$^{-3}$ and the vertical force $|K_{z=1.1}|/2\pi G=77 M_\odot $~\rm pc$^{-2}$. A comparative analysis of the refined potential models is made and for each of the models the estimates of a number of the Galactic characteristics are presented.}

\end{abstract}
  \keywords{Galaxy: kinematics and dynamics, structure}
 % \classification[PACS]{}
 % \communicated{...}
 % \dedication{...}

  \journalname{Open Astronomy}
  \DOI{DOI}
  \startpage{1}
  \received{..}
  \revised{..}
  \accepted{..}
  \journalyear{2017}
  \journalvolume{1}
%  \journalissue{1}

\maketitle
\section{Introduction}
The refinement of the Galactic gravitational potential model is an important task of stellar astronomy. To solve this problem, in the first place, we need high-precision measurement data on distances and velocities of a large number of Galactic objects. In this respect,  the compilation of Bhattacharjee et al. (2014) is of great interest. In the compilation, practically all known measurements of Galactic objects (Hydrogen clouds, maser sources, large sample of individual stars, globular clusters and dwarf-galaxy companions of the Milky Way), located in a wide range of Galactocentric distances $ R: 0-200 $~kpc, are presented. This paper is based on the use of these data in the task of construction of the Galactic rotation curve.

To date, there exists a large number of different models of the Galactic gravitational potential, described by analytic expressions. As a rule, these are multi-component (sometimes up to six components) models that describe
contributions of:

(i) the central condensation or bulge of the Galaxy,

(ii) the Galactic disk, which is sometimes represented as a combination
of several components (thin and thick discs, stellar and gas components),and

(iii) a halo of invisible matter that dominates on the large distances (approximately
$ R> 30 $ ~ kpc), and gives the largest contribution to the total Galactic mass.

Additional constraints, such as values of the local dynamical matter
density $\rho_\odot$ and vertical force $|K_z|$ are important for the construction of an
adequate model of the Galaxy from a physical point of view.

Let us note that, even with the use of modern high-precision measurement data, the estimates of the Galactic mass can differ in 2--3 times. For example, in a paper by Watkins et al. (2010) the most probable value of the Galactic mass within a sphere of radius 300 kpc was evaluated as $ M_{300}=(0.9\pm0.3)\times10^{12}M_\odot$ for the case of isotropic distribution of velocities of the halo stars. However, as these authors note, the mass estimate is sensitive to the assumed anisotropy and could plausibly lie between $ (0.7-3.4)\times10^{12}M_\odot. $
From the analysis of the motion of dwarf-galaxy companions of the Milky Way it is known that the inclusion or non-inclusion in the sample of one of the most distant supposed companions of the Galaxy, Leo ~ I
(McConnachie 2012), changes the estimate of the Galactic mass in 2-3 times (Watkins et al. 2010; Sohn et al. 2013; Boylan-Kolchin et al. 2013; Bajkova, Bobylev 2017).

In this paper, the results of recent works of Bobylev, Bajkova (2013), Bajkova, Bobylev (2016) and Bobylev et al. (2017) are summarized. These works are devoted to the refinement of the parameters of six selected, most popular axisymmetric three-component Galactic potential models, differing by expressions for the description of the Galactic halo. These are the models of: Allen-Santill\'an (I), Wilkinson-Evans (II), Navarro-Frenk-White (III), Binney (IV), Plummer (V), and Hernquist (VI). The bulge and disk components in all the models are described by the Miyamoto-Nagai expressions. In this paper, an analytical review of the obtained models and the corresponding estimates of a number of physical characteristics of the Galaxy are given.

The work is structured as follows. Section 2 lists the measurement data used. Section 3 provides an overview of all the necessary analytical expressions for implementation of fitting operations, as well as calculation of main characteristics of the Galaxy. Section 4 gives an overview of the results of fitting, as well as a comparative analysis of the obtained model rotation curves and other Galactic characteristics. Section 5 devoted to discussion of the results. Section 6 presents concluding remarks.

\section{Data}
The data used include:

(i)the line-of-sight velocities of HI clouds at the tangent points from Burton, Gordon (1978). These data on the rotation curve fill the range of distances $R<4$~kpc;

(ii)a sample of masers with measured trigonometric parallaxes, proper motions, and line-of-sight velocities (Reid et al. 2014). They are located in the interval of distances $R$ from 4 to 20 kpc;

(iii)the average circular rotation velocities from the work of Bhattacharjee et al. (2014) calculated using objects at distances $R$ from 20 to $\approx$200 kpc. These are the velocities of 1457 blue horizontal branch
giants, 2227 K giants, 16 globular clusters, 28 distant halo giants, and 21 dwarf galaxies.

Since Bhattacharjee et al. (2014) constructed the Galactic rotation curve with $R_\odot=8.3$~kpc and $V_\odot=244$ km s$^{-1}$, we also calculate the model circular velocities of objects with these parameters.

\section{Potential Models}
In all of the models here, the axisymmetric Galactic potential is represented as a sum of three components --- a central spherical bulge $\Phi_b(r(R,z))$, a disk $\Phi_d(r(R,z))$, and a massive spherical
dark matter halo $\Phi_h(r(R,z))$:
 \begin{equation}
 \begin{array}{lll}
 \renewcommand{\arraystretch}{2.8}
  \Phi(R,z)=\Phi_b(r(R,z))+\Phi_d(r(R,z))+\\
  \qquad+\Phi_h(r(R,z)).
 \label{pot}
 \end{array}
\end{equation}
We use a cylindrical coordinate system ($R,\psi,z$) with the coordinate origin at the Galactic center. In a rectangular coordinate system $(x,y,z)$ with the coordinate origin at the Galactic center, the distance to a star (spherical radius) will be $r^2=x^2+y^2+z^2=R^2+z^2$.

In accordance with the convention adopted in Allen and Santill\'an (1991), we express the gravitational potential in units of 100 km$^2$~s$^{-2}$, the distances in kpc, and the masses in units of the Galactic mass $M_{gal}=2.325\times 10^7 M_\odot$, corresponding to the gravitational constant $G=1.$

The expression for the mass density follows from the Poisson equation
\begin{equation}
4\pi G\rho(R,z)=\nabla^2\Phi(R,z) \label{pois1}
\end{equation}
and is
\begin{equation}
 \rho(R,z)=\frac{1}{4\pi G}\Bigg{(}
 \frac{\partial^2\Phi(R,z)}{\partial R^2}+\frac{\partial\Phi(R,z)}{R \partial R}+
 \frac{\partial^2\Phi(R,z)}{\partial z^2}\Bigg{)}.
\label{pois2}
\end{equation}
The force (vertical force) acting in the $z$ direction perpendicularly to the Galactic plane is expressed as
\begin{equation}
K_z(z,R)=-\frac{\partial\Phi(z,R)}{\partial z}. \label{Kz}
\end{equation}
Eqs.~(\ref{pois2}) and (\ref{Kz}) are needed to solve the problem of fitting the parameters of the gravitational potential models with constraints imposed on the local dynamical mass density
$\rho_\odot$ and the force $|K_z(z,R_\odot)|$ at $z=1.1$~kpc, which are known from observations. In addition, the following expressions are needed to calculate:

1) the circular velocities
\begin{equation}
V_{circ}(R)=\sqrt{R\frac{d\Phi(R,0)}{dR}}, \label{V}
\end{equation}

2) the Galactic mass contained in a sphere of radius $r$
\begin{equation}
m(<r)=4\pi \int_0^r RdR \int_0^{\sqrt{r^2-R^2}}\rho(R,z)dz, \label{m}
\end{equation}

3) the parabolic velocity or the escape velocity of a star from the attractive Galactic field
\begin{equation}
V_{esc}(R,z)=\sqrt{-2\Phi(R,z)}, \label{Vesc}
\end{equation}

4) the Oort parameters
\begin{equation}
A=-\frac{1}{2}R_\odot\Omega_\odot^{'}, \qquad B=A-\Omega_\odot,
\label{AB}
\end{equation}
where $\Omega=V/R$ is the angular velocity of Galactic rotation $(\Omega_\odot=V_\odot/R_\odot)$, $\Omega^{'}$ is the first derivative of the angular velocity with respect to $R,$ and $R_\odot$ is the Galactocentric distance of the Sun.

5) the surface density of gravitating matter within $z_{out}$ of the Galactic $z=0$ plane
 \begin{equation}
 \begin{array}{lll}
 \renewcommand{\arraystretch}{2.8}
 \displaystyle
 \Sigma_{out}(z_{out})=  2\int_0^{z_{out}} \rho(R,z)dz=\\
   \qquad \displaystyle =\frac{K_{z}}{2\pi G}+\frac{2z_{out}(B^2-A^2)}{2\pi G},
 \label{Sigma}
 \end{array}
\end{equation}
where $K_{z}$ corresponds to $z_{out}$.

In all of the models being considered here, the bulge, $\Phi_b(r(R,z))$, and disk, $\Phi_d(r(R,z))$, potentials are represented in the form proposed by Miyamoto and Nagai (1975):
 \begin{equation}
  \Phi_b(r)=-\frac{M_b}{(r^2+b_b^2)^{1/2}},
  \label{bulge}
 \end{equation}
 \begin{equation}
 \Phi_d(R,z)=-\frac{M_d}{\{R^2+[a_d+(z^2+b_d^2)^{1/2}]^2\}^{1/2}},
 \label{disk}
\end{equation}
where $M_b$ and $M_d$ are the masses of the components, $b_b, a_d,$ and $b_d$ are the scale lengths of the components in kpc.

Below we give expressions for the six dark matter halo potential models.

{\bf Model I.} The expression for the halo potential was derived by Irrgang et al. (2013) from the expression for the halo mass (Allen, Martos, 1986). It slightly differs from that given in Allen and Santill\'an
(1991) and is
 \begin{equation}
 \renewcommand{\arraystretch}{2.8}
  \Phi_h(r) = \left\{
  \begin{array}{ll}
  \displaystyle
  \frac{M_h}{a_h}\biggl( \frac{1}{(\gamma-1)}\ln
  \biggl(\frac{1+(r/a_h)^{\gamma-1}}{1+(\Lambda/a_h)^{\gamma-1}}\biggr)-\\
    \displaystyle
    \qquad\qquad-\frac{(\Lambda/a_h)^{\gamma-1}}{1+(\Lambda/a_h)^{\gamma-1}}\biggr),
  \quad\textrm{if }   r\leq \Lambda & \\
  \displaystyle
  -\frac{M_h}{r} \frac{(\Lambda/a_h)^{\gamma}}{1+(\Lambda/a_h)^{\gamma-1}},
  \qquad\qquad\textrm{if }  r>\Lambda, &
  \end{array}\right.
 \label{halo-I}
 \end{equation}
where $M_h$ is the mass, $a_h$ is the scale length, the Galactocentric distance is $\Lambda=200$ kpc, and the
dimensionless coefficient $\gamma=2.0$.

{\bf Model II.} The halo component is represented in the form proposed by Wilkinson, Evans (1999) as
 \begin{equation}
  \Phi_h(r)=-\frac{M_h}{a_h} \ln {\Biggl(\frac{a_h+\sqrt{r^2+a^2_h}} {r}\Biggr)}.
 \label{halo-II}
 \end{equation}

{\bf Model III.} The halo component is represented in the form proposed by Navarro et al. (1997) as
 \begin{equation}
  \Phi_h(r)=-\frac{M_h}{r} \ln {\Biggl(1+\frac{r}{a_h}\Biggr)}.
 \label{halo-III}
 \end{equation}
This model is often called the NFW (Navarro--Frenk--White) model.

{\bf Model IV.} The halo component is represented by a logarithmic potential in the form proposed by Binney (1981):
 \begin{equation}
 \renewcommand{\arraystretch}{1.2}
  \Phi_h(R,z)=-\frac{v^2_0}{2}\ln\Biggl(R^2+a^2_h+\frac{z^2}{q^2_{\Phi}}\Biggr),
 \label{halo-IV}
 \end{equation}
where $v_0$ is the velocity in km s$^{-1}$, $q_\Phi$ is the axial ratio of the ellipsoid: $q_\Phi=1$ for a spherical halo, $q_\Phi<1$ for an oblate one, and $q_\Phi>1$ for a prolate one. We take $q_\Phi=1.$

{\bf Model V.} In this model we use a Plummer (1911) sphere (coincident with Eq.~(\ref{bulge})) to describe the halo potential:
 \begin{equation}
  \Phi_h(r)=-\frac{M_h}{(r^2+a_h^2)^{1/2}}.
  \label{halo-V}
 \end{equation}

{\bf Model VI.} The halo component is represented by the Hernquist (1990) potential, which
is a special case of the formula proposed by Kuzmin, Veltmann (1973):
 \begin{equation}
  \Phi_h(r)=-\frac{M_h}{r+a_h}.
  \label{halo-VI}
 \end{equation}

\subsection{Parameter Fitting}
The parameters of the potential models are found by least-squares fitting to the measured rotation velocities of the Galactic objects. We use the unit weights, because they provide the smallest absolute residual between the data and the model rotation curve. In addition, we used (Irrgang et al. 2013) the constraints on (i)
the local dynamical matter density $\rho_\odot=0.1M_\odot$~pc$^{-3}$ and (ii) the force acting perpendicularly to the Galactic plane or, more specifically, $|K_{z=1.1}|/2\pi G=77M_\odot$~pc$^{-2}$.

The local dynamical matter density $\rho_\odot$, which is the sum of the bulge, disk, and dark matter densities in a small solar neighborhood, together with the surface density $\Sigma_{1.1}$ are the most important additional constraints in the problem of fitting the parameters of the potential models to the measured circular velocities (Irrgang et al. 2013):
\begin{equation}
\rho_\odot=\rho_b(R_\odot)+\rho_d(R_\odot)+\rho_h(R_\odot),
\label{ro}
\end{equation}
\begin{equation}
\Sigma_{1.1}=
  \int\limits^{1.1\,\hbox {\footnotesize\it kpc}}_{-1.1\,\hbox {\footnotesize\it kpc}}
(\rho_b(R_\odot,z)+\rho_d(R_\odot,z)+\rho_h(R_\odot,z))dz.
\label{Sig}
\end{equation}
The surface density is closely related to the force $K_z(z,R)$ in accordance with Eq.~(\ref{Sigma}). Since the two most important parameters $\rho_\odot$ and $|K_z|/2\pi G$ are known from observations with a sufficiently high accuracy, introducing additional constraints on these two parameters allows the parameters of the gravitational
potential to be refined significantly.

As a result, the parameter fitting problem was reduced to minimizing the following quadratic functional $F:$
\begin{equation}
 \begin{array}{lll}
 \renewcommand{\arraystretch}{2.8}
 \displaystyle
 \min F=\sum_{i=1}^N
 (V_{circ}(R_i)-\widetilde{V}_{circ}(R_i))^2+\\
 +\alpha_1(\rho_\odot-\widetilde{\rho}_\odot)^2+\alpha_2(|K_{z=1.1}|/2\pi G %-\\
 -|\widetilde{K}_{z=1.1}|/2\pi G)^2,
  \label{F}
 \end{array}
\end{equation}
where the measured quantities are denoted by the tilde, $R_i$ are the distances of the objects with measured circular velocities, $\alpha_1$ and $\alpha_2$ are the weight factors at the additional constraints that were chosen so as to minimize the residual between the data and the model rotation curve provided that the
additional constraints hold with an accuracy of at least 5\%. Based on the constructed models, we calculated the local surface density of the entire matter $\rho_\odot$ and $|K_{z=1.1}|/2\pi G$ related to $\Sigma_{1.1}$ and $\Sigma_{out}.$ The errors of all the parameters given in Tables~\ref{t:01}--\ref{t:02} were determined through Monte Carlo simulations.

\begin{figure}[t]
\begin{center}
\includegraphics[width=68mm]{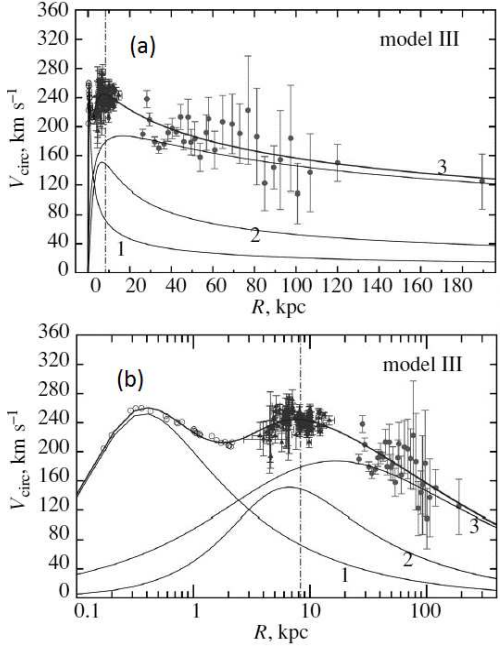}
\caption{Galactic rotation curve for model III in linear (a) and
logarithmic (b) distance scales; the vertical line marks the Sun's
position, numbers 1, 2, and 3 denote the bulge, disk, and halo
contributions, respectively; the open circles, filled triangles,
and filled circles indicate the HI velocities, the velocities of
masers with measured trigonometric parallaxes, and the velocities
from Bhattacharjee et al. (2014), respectively. }
 \label{f1}
\end{center}  \end{figure}

%%%%%%%%%%%%%%%%%%%%%%%%%%%%%%%%%%%%%%%%%%%%%%%%
 {\begin{table*}[t]                            %% t~1.
 \caption[] {\small\baselineskip=1.0ex
 The parameters of models I--VI found by fitting to the observational data}
 \label{t:01}
 \small \begin{center}\begin{tabular}{|r|r|r|r|r|r|r|}
       Parameters &  Model~I&  Model~II &  Model~III & Model~IV & Model~V  & Model VI \\\hline
  $M_b$ ($M_{g}$) & 386 $\pm$10 & 142$\pm$12    &  443$\pm$27   & 486$\pm$10 &456$\pm$40 & 461$\pm$22 \\
  $M_d$ ($M_{g}$) & 3092$\pm$62 & 2732$\pm$16   &  2798$\pm$84  & 3079$\pm$23 & 3468$\pm$71 &2950$\pm$33\\
  $M_h$ ($M_{g}$) &  452$\pm$83 &24572$\pm$5459 & 12474$\pm$3289& $^*$14210$\pm$858 & 16438$\pm$1886&29677$\pm$2791\\
  $b_b$ (kpc)     &0.249$\pm$0.006&0.250$\pm$0.009&0.267$\pm$0.009&0.277$\pm$0.005&0.265$\pm$0.006&0.272$\pm$0.013 \\
  $a_d$ (kpc)     &3.67 $\pm$0.16 & 5.16$\pm$0.32& 4.40$\pm$0.73&3.54$\pm$0.06&2.94$\pm$0.076&3.85$\pm$0.08   \\
  $b_d$ (kpc)     &0.305$\pm$0.003&0.311$\pm$0.003&0.308$\pm$0.005&0.300$\pm$0.002&0.313$\pm$0.002 & 0.309$\pm$0.001\\
  $a_h$ (kpc)     &1.52 $\pm$0.18 &64.3 $\pm$15 & 7.7$\pm$2.1&3.20$\pm$0.45&16.57$\pm$1.38& 21.27$\pm$1.06  \\\hline
    Entropy $E$   &  -31.40 & -27.78 & -24.51&  -29.11   &  -29.72  &  -24.96 \\\hline
  $\delta $ (km/s)      &     15.7  &   13.8   &   13.1 &     15.04 &   14.89  &   13.23   \\\hline
  $\delta_{irg}$ (km/s) &     19.4  &   16.4   &   38.4 &  - &  - &  -   \\
 \end{tabular}\end{center}
  {\small Note:
  1)~The Galactic mass unit is $M_{g}=2.325\times10^7 M_\odot$,
  2)$^*$ $v_0^2/2$ in km$^2$s$^{-2}$ is given here.
  }
 \end{table*}}
%%%%%%%%%%%%%%%%%%%%%%%%%%%%%%
 {\begin{table*}[t]                            %% t~2.
 \caption[] {\small\baselineskip=1.0ex
 The Galactic physical characteristics calculated from the parameters of models I--VI }
 \label{t:02}
 \small \begin{center}\begin{tabular}{|r|r|r|r|r|r|r|}
           Parameters &         Model~I &        Model~II &       Model~III &        Model~IV & Model~V & Model~VI \\\hline
$(\rho_\odot)_d$      & 0.092$\pm$0.010 & 0.090$\pm$0.010 & 0.089$\pm$0.011 & 0.092$\pm$0.009 &0.089$\pm$0.010& 0.090$\pm$0.010\\
$(\rho_\odot)_h$      & 0.008$\pm$0.001 & 0.010$\pm$0.001 & 0.010$\pm$0.001 & 0.008$\pm$0.001 &0.011$\pm$0.001& 0.011$\pm$0.001\\
    $\rho_\odot$      & 0.100$\pm$0.010 & 0.100$\pm$0.010 & 0.100$\pm$0.010 & 0.100$\pm$0.010 &0.100$\pm$0.010&0.100$\pm$0.010  \\
 $|K_{z=1.1}|/2\pi G$   &  77.2$\pm$6.9   & 77.01$\pm$10.2  &  77.1$\pm$12.5  &  77.0$\pm$6.3   & 77.1$\pm$6.6& 77.2$\pm$5.8  \\
 $\Sigma_{1.1}$       &  71.4$\pm$7.3   & 75.78$\pm$10.1  &  76.8$\pm$12.3  &  71.4$\pm$6.4   & 78.6$\pm$7.9    & 76.9$\pm$6.4  \\
 $\Sigma_{out}$       & 44.73$\pm$8.25  &  66.7$\pm$10.0  &  69.9$\pm$17.6  &  45.2$\pm$7.1   & 75.0$\pm$14.2   & 68.9$\pm$10.1 \\
 $V_{esc,R=R_\odot}$  & 561.4$\pm$46.5  & 518.0$\pm$56.2  & 537.8$\pm$70.1  & 450.2$\pm$8.6   &516.0$\pm$21.4&524.8$\pm$18.2 \\
 $V_{esc,R=200\,kpc}$ & 250.0$\pm$25.6  & 164.4$\pm$16.0  & 210.6$\pm$26.2  & 550.7$\pm$16.7  &142.5$\pm$5.7& 173.9$\pm$6.8\\
 $V_\odot$            & 239.0$\pm$12.0  & 242.5$\pm$28.0  & 243.9$\pm$34.5  & 241.3$\pm$3.9  & 238.8$\pm$9.4   & 243.1$\pm$6.8 \\
      $A$             & 16.01$\pm$0.80  & 15.11$\pm$1.84  & 15.04$\pm$2.37  & 16.10$\pm$0.62  &  14.49$\pm$0.60 & 15.05$\pm$0.52 \\
      $B$             &-12.79$\pm$1.06  &-14.10$\pm$1.77  &-14.35$\pm$2.12  &-12.97$\pm$0.69  & -14.27$\pm$1.15 & -14.24$\pm$0.84 \\
 $M_{~50\,kpc}$       & 0.415$\pm$0.074 & 0.416$\pm$0.094 & 0.406$\pm$0.115 & 0.409$\pm$0.020 & 0.417$\pm$0.034 & 0.417$\pm$0.032 \\
 $M_{100\,kpc}$       & 0.760$\pm$0.149 & 0.546$\pm$0.108 & 0.570$\pm$0.153 & 0.738$\pm$0.040 & 0.457$\pm$0.037 & 0.547$\pm$0.042 \\
 $M_{150\,kpc}$       & 1.105$\pm$0.224 & 0.591$\pm$0.114 & 0.674$\pm$0.177 & 1.066$\pm$0.061 & 0.466$\pm$0.037 & 0.607$\pm$0.047\\
 $M_{200\,kpc}$       & 1.450$\pm$0.300 & 0.609$\pm$0.117 & 0.750$\pm$0.194 & 1.395$\pm$0.082 & 0.469$\pm$0.038 & 0.641$\pm$0.049\\
  \end{tabular}\end{center}
  {\small Note: $\rho_\odot$ in $M_\odot$ pc$^{-3}$, $|K_z|/2\pi G$ in $M_\odot$ pc$^{-2}$,
  $\Sigma$ in $M_\odot$ pc$^{-2}$, $V_{esc}$, $V_\odot$ in km s$^{-1}$,
  $A$ and $B$ in km s$^{-1}$ kpc$^{-1}$, $M_G$ in $10^{12}M_\odot$ are given here.}
 \end{table*}}

 \section{Overview of fitting results}
Table~\ref{t:01} brings together the values of the seven parameters ($M_b, M_d, M_h, b_b, a_d, a_b, a_h$), which were found by solving the fitting problem for the six Galactic potential models(Bajkova, Bobylev 2016;
Bobylev et al. 2017). The value $\delta$ in the table gives the residuals (in km s$^{-1}$) between the model rotation curve found and the circular velocities

 $\delta^2=\biggl(\sum_{i=1}^N(V_{circ}(R_i)-\widetilde{V}_{circ}(R_i))^2\biggr)/N$.

To estimate the degree of uniformity of the residual noise (the difference between the data and the model rotation curve), we used the well-known concept of entropy for bipolar signals (Bajkova 1992) calculated as follows:
$$
E=-\frac{1}{N}\sum_{i=1}^N |\Delta_i|\ln(|\Delta_i|),
$$
where $\Delta_i=V_{circ}(R_i)-\widetilde{V}_{circ}(R_i)$. The higher the entropy, the more uniform the noise and, consequently, the better the parameter fitting. Obviously, the combination of $\delta$ and $E$ gives a more comprehensive idea of the quality of fitting by various models than does $\delta$ alone. The entropy of
the residual noise is given in Table~\ref{t:01} as well.

As can be seen, model III provides the best fit to the data, model VI yields a comparable result. I.e., model III provides the smallest residual $\delta$ and the greatest entropy of the noise, i.e., its uniformity. Therefore we consider model III as the best among others under consideration.

For comparison, the last row in table~\ref{t:01} gives the residuals between our data and the model rotation curves from Irrgang et al. (2013). It can be seen that the parameters we found, provide a more accurate fit, especially in the case of model III (we managed to reduce the residual by a factor of 3).

Table~\ref{t:02} gives the physical quantities calculated from the derived parameters of the potential models (Eqs.~(\ref{pois2})--(\ref{Sigma})). These include the local disk density $(\rho_\odot)_d$ (the local bulge density is not given, because it is lower than the local disk density by several orders of magnitude), the local dark matter density $(\rho_\odot)_h,$ the local density of the entire matter $\rho_\odot,$ the local surface
density $\Sigma_{1.1}$ and $\sum_{out}$, the two escape velocities from the Galaxy $V_{esc}$ (\ref{Vesc}) for $R=R_\odot$ and $R=200$~kpc, the linear circular rotation velocity of the Sun $V_\odot,$ the Oort constants $A$ and $B$ from Eqs.~(\ref{AB}), and the Galactic mass $M_G$ for four radii of the enclosing sphere. As an example, the best model Galactic rotation curve corresponding to potential model III is presented in Figure~\ref{f1}. All the rotation curves constructed for all the  potential models are given in papers by Bajkova, Bobylev (2016) and Bobylev et al. (2017).

Let us perform a comparative analysis of the constructed model rotation curves.

In model I, the function describing the halo contribution to the velocity curve is a nondecreasing one. For this reason, the resulting model rotation curve describes poorly the data already at distances $R$ greater than 120 kpc, the Galactic mass estimate at $R\leq200$~kpc is the greatest compared to the remaining models in this paper.

As can be seen from Table~\ref{t:01}, the lowest-mass central component corresponds to model II. We consider that although model II describes satisfactorily the Galactic rotation curve in the $R$ range 0--200~kpc, it suggests the presence of a substantial dark matter mass in the inner region of the Galaxy, which is most likely far from reality.

Model III is currently one of the most commonly used models (see, e.g., Sofue 2009; Kafle et al. 2012; Deason et al. 2012a). In the outer Galaxy ($R>R_\odot$) its properties are similar to those of model II, while in the inner Galaxy ($R<R_\odot$) the dark matter mass is insignificant, which favourably distinguishes this model from model II. As can be seen from the next-to-last row in Table~\ref{t:01}, this model fits the data with the smallest
residual $\delta$ and the greatest entropy of the residual noise.

In model IV, in accordance with (\ref{V}), the circular velocity of the halo increases monotonically with Galactocentric distance.  In this model, it is apparently desirable to artificially correct the halo density function at great distances ($R>200$~kpc), as is done in model I.

Model V has the largest disk mass ($M_d$) compared to our other models, as can be seen from Table~\ref{t:01}. It follows from the last rows in Table~\ref{t:02}, that based on this model, we obtain the smallest Galactic mass ($M_G$) among the other models. Models V and VI are attractive in that both the circular velocity of the
halo and the overall rotation curve at distances greater than 100 kpc fall off gently. Therefore, there is no need to artificially correct the halo density function.

So, the Galactic mass within a sphere of radius 50 kpc, $M_{50}\approx(0.41\pm0.12)\times10^{12}M_\odot$,
was shown to satisfy all six models. The differences between the models become increasingly significant with increasing radius $R.$ In model I, the Galactic mass within a sphere of radius 200 kpc
turns out to be greatest among the models considered,
 $M_{200}=(1.45\pm0.30)\times10^{12}M_\odot$,
the smallest value was found in model V,
 $M_{200}=(0.469\pm0.038)\times10^{12}M_\odot.$
Model III (Navarro et al. 1997), which is the best one among
those considered ensures the Galactic mass
 $M_{200}=(0.75\pm0.19)\times10^{12}M_\odot.$
The model VI is close to the model III with
 $M_{200}=(0.64\pm0.05)\times10^{12}M_\odot$.

Such local parameters of the rotation curve as the velocity $V_\odot$ and the Oort constants $A$ and $B$ are well reproduced by the models considered. In model V, however, $B$ is comparable in absolute value to $A.$ Therefore, the rotation velocity in a small segment near the Sun is nearly flat ($V_{circ}=const$).

Interestingly, the escape velocity $V_{esc}$ ($R=200$~kpc) is usually approximately half that at $R=R_\odot.$ However, for model IV the parabolic velocity at $R=200$~kpc exceeds its value calculated for $R=R_\odot.$

 \begin{figure}[t]
 \begin{center}
 \includegraphics[width=78mm]{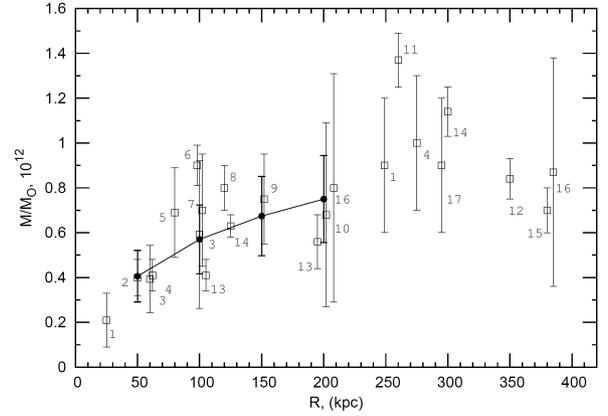}
 \caption{The Galactic mass estimates obtained by various authors (open
squares) and the estimates found in this paper based on model III
(thick line); the numbers indicate the following sources: 1--Kafle
et al. (2012), 2--Deason et al. (2012a), 3--Bhattacharjee et al.
(2013), 4--Xue et al. (2008), 5--Gnedin et al. (2010), 6--McMillan
(2011), 7--Dehnen and Binney (1998), 8--Battaglia et al. (2005),
9--Deason et al. (2012b), 10--Bhattacharjee et al. (2014),
11--Eadie et al. (2015), 12--Karachentsev et al. (2009),
 13--Gibbons et al. (2014), 14--Eadie et al. (2017), 15--Sofue (2012),
 16--Sofue (2015), 17--Watkins et al. (2010).
 }
 \label{f2} \end{center}  \end{figure}

 \section{Discussion}
It is important to note that to determine an adequate Galactic rotation curve it is highly desirable to have observational data covering a wide range of Galactocentric distances $R$. For example, in Irrgang et al. (2013) the construction of potential models was based on the high-precision measurements of masers, but located no
farther than 20 kpc from the Galactic center. For model III we found
 $M_{200}=(0.75\pm0.19)\times10^{12}M_\odot,$
 while Irrgang et al. (2013), based on extrapolation, estimated
  $M_{200}=(3.0\pm1.1)\times10^{12}M_\odot.$ Thus, here we have a significant discrepancy.

Therefore, it is of interest to compare our Galactic mass estimates with the results of other authors obtained from objects far from the Galactic center.

There is a vast literature on this issue. We note, for example, the work of Carlesi et al. (2017), which contains a large summary of modern estimates of the Galactic mass. True, it contains a rather large number of estimates of the virial mass of the Galaxy without indicating the exact value of the virial radius.

Figure~\ref{f2} presents the selected results of various authors, obtained by independent methods. Compared to Figure~7 from work of  Bajkova, Bobylev (2016), here a number of other results is added. Note that the results marked by numbers 1 and 4 at $R>250$~kpc are the virial mass estimates, while the direct estimates 1 and 4 were obtained from the data at $R<80$~kpc.

Xue et al. (2008) analysed the line-of-sight velocities of blue horizontal-branch giants at distances $R<60$~kpc. They constructed a three-component potential model in which the dark halo mass was represented in the NFW form, while the bulge and disk potentials differ from those we used.

In the works of Eadie et al. (2015, 2017), Watkins et al. (2010), and as well as Sofue (2009; 2015), for estimating the Galactic mass data on globular clusters and dwarf galaxies were used. The analysis method applied in the works of Sofue (2009; 2015) is close to ours: there was constructed the Galactic rotation curve, then was improved the model of the Galactic gravitational potential, and finally the latter was used  for the Galactic mass estimation.
Eadie et al. (2017) present a hierarchical Bayesian method, which uses a distribution function to model the Galaxy and kinematic data from companion objects, such as globular clusters, to trace the Galaxy's gravitational potential.

Interesting estimates were obtained in work of Gibbons et al. (2014) from analysis of observations of the Sagittarius stream. This stream is a tail, formed as a result of the destruction of the dwarf
galaxy after several turns around the Galactic center. The presence of such a loop, allows just to see the orbit
of the Galaxy companion, which ultimately makes it possible to clarify the gravitational potential of the Milky Way.

At large distances, $R>50$~kpc, and not for all objects, a reliable estimate of circular rotational velocities
$ V_{circ} $ is achieved by the direct methods. Quite often it is used an indirect method of determining
such velocities, based on the Jeans equation (Binney, Tremaine, 1987). This equation allows to estimate the velocities $ V_{circ} $ through observed dispersions of radial velocities $ \sigma_r $. The data, obtained in such way, were used, for example, in the works of Battaglia et al. (2005), Gnedin et al. (2010) and Bhattacharjee et al. (2014).

In the work of Karachentsev et al. (2009) for estimating the mass of the Galaxy an independent method based on the effect of local Hubble flow deceleration was used. For the analysis, the line-of-sight velocities of a large number of dwarf galaxies of a Local group was used.

In general, we can conclude that there is a good agreement between our results based on model III (and the model VI close to it) and results of other authors. And, as you can see from Figure~\ref{f2}, good agreement within the available errors can be traced (if you continue mentally the curve) to the largest values of $R.$ At the same time, if you chart in Figure~\ref{f2} the found mass values for our model I, then at distances $ R\geq150 $ ~ kpc they will differ significantly from the results of other authors.

\section{Conclusion}
Thus, in this paper we present the refined parameters of six most popular Galactic gravitational potential models differing by the shape of the dark matter halo (Allen-Santill\'an (I), Wilkinson-Evans (II), Navarro-Frenk-White (III), Binney (IV), Plummer (V), and Hernquist (VI)). In all the models considered, the central component (bulge) and the Galactic disk are represented in the form of Miyamoto and Nagai (1975). New parameters are obtained by fitting to modern data on the circular rotation velocities of Galactic objects covering a wide range of Galactocentric distances (up to 200 kpc), as well as accounting for the restrictions on the local parameters of the Galaxy (local dynamical matter density and vertical force). The best in terms of accuracy of fit is model III. The model VI is the closest to model III. Model I shows the worst result. At distances up to 50 kpc from the Galactic center, all models provide approximately equal physical characteristics. So, the mass of the Galaxy inside a sphere of radius 50 kpc, for all models is
 $M_{50}\approx(0.41\pm0.12)\times 10^{12}M_\odot$.
The difference between the models increases with increasing radius $ R. $
In model I, the Galactic mass within a sphere of radius 200 kpc turns out to be greatest among the models considered, $M_{200}=(1.45\pm0.30)\times10^{12}M_\odot$, the smallest value was found in model V,
 $M_{200}=(0.469\pm0.038)\times10^{12}M_\odot.$

In our view, model III (NFW) is the best one among those considered, because it ensures the smallest residual between the data and the constructed model rotation curve provided that the constraints on the local
parameters hold with a high accuracy. The model VI is the closest to the model III. In models III and VI the Galactic mass is $M_{200}=(0.75\pm0.19)\times10^{12}M_\odot$ and $M_{200}=(0.64\pm0.05)\times10^{12}M_\odot$ respectively. We have shown, that there is a good agreement between our estimates based on model III and the results of other authors obtained by independent methods.

\section{Acknowledgment}
We are grateful to the referee for the remarks that contributed to an improvement of the paper.
This work was supported by the ``Transient and Explosive Processes in Astrophysics'' Program P--7 of the Presidium of the Russian Academy of Sciences.

\section{References}
\begin{itemize}

 \item Allen, C., Martos, M. 1986, Revista Mexicana Astron. Astrof., 13, 137-147

 \item Allen, C., Santill\'an, A. 1991, Revista Mexicana Astron. Astrofis., 22, 255-263

 \item Bajkova, A.T. 1992, A\&AT, 1, 313-320

 \item Bajkova, A.T., Bobylev, V.V. 2016,  Astron. Lett. 42, 567-582 %Pot-I

 \item Bajkova, A.T., Bobylev, V.V., 2017,  Astron. Rep. 61, 727-738

 \item Battaglia, G., Helmi, A., Morrison, H., et al. 2005, MNRAS, 364, 433-442

 \item Bhattacharjee, P., Chaudhury, S., Kundu, S., et al. 2013, Phys. Rev. D, 87, id. 083525

 \item Bhattacharjee, P., Chaudhury, S., Kundu, S. 2014, ApJ, 785, 63, 13pp

 \item Binney, J. 1981, MNRAS, 196, 455-467

 \item Binney J., Tremaine S., 1987, Galactic Dynamics. Princeton Univ. Press, Princeton, NJ

 \item Bobylev, V.V., Bajkova, A.T. 2013,  Astron. Lett. 39, 809-818

 \item Bobylev, V.V., Bajkova, A.T., Gromov, A.O. 2017,  Astron. Lett. 43, 241-251 %Pot-II

 \item Boylan-Kolchin, M., Bullock J. S., Sohn S. T., et al. 2013, ApJ, 768, 140, 11pp

 \item Burton, W.B., Gordon, M.A. 1978, A\&A, 63, 7-27

 \item Carlesi, E., Hoffman, Y., Sorce, J.G., et al. 2017, MNRAS, 465, 4886-4894

 \item Deason, A.J., Belokurov, V., Evans, N.W., et al. 2012a, MNRAS, 424, L44-L48

 \item Deason, A.J., Belokurov, V., Evans, N.W., et al. 2012b, MNRAS, 425, 2840-2853

 \item Dehnen, W., Binney, J. 1998, MNRAS, 294, 429-438

 \item Eadie, G.M., Harris, W.E., Widrow, L.M. 2015, ApJ, 806, 54, 14pp

 \item Eadie, G.M., Springford, A., Harris, W.E., 2017, ApJ, 835, 167, 9pp

 \item Gibbons, S., Belokurov, V., Evans, N.W. 2014, MNRAS, 445, 3788-3802

 \item Gnedin, O.Y., Brown, W.R., Geller, M.J., et al. 2010, ApJ, 720, L108-L112

 \item Hernquist, L. 1990, ApJ, 356, 359-364

 \item Irrgang, A., Wilcox, B., Tucker, E., et al. 2013, A\&A, 549, A137, 13pp

 \item Kafle, R.R., Sharma, S., Lewis, G.F., et al. 2012, ApJ, 761, 98, 17pp

 \item Karachentsev, I.D., Kashibadze, O.G., Makarov, D.I., et al. 2009, MNRAS, 393, 1265-1274

 \item Kuzmin, G.G., Veltmann, \"{U}.-I.K. 1973, Publ. Tartu Astrofiz. Obs., 40, 281-323

 \item McConnachie, A.W. 2012, AJ, 144, 4, 36pp

 \item McMillan, P.J. 2011, MNRAS, 414, 2446-2457

 \item Miyamoto M., Nagai, R. 1975, PASJ, 27, 533-543

 \item Navarro, J.F., Frenk, C.S., White, S.D.M. 1997, ApJ, 490, 493-508

 \item Plummer, H.C. 1911, MNRAS, 71, 460-470

 \item Reid, M.J., Menten, K.M., Brunthaler, A., et al. 2014, ApJ, 783, 130, 14pp

 \item Sofue, Y. 2009, PASJ, 61, 153-161

 \item Sofue, Y. 2012,  PASJ, 64, 75, 8pp

 \item Sofue, Y. 2015, PASJ, 67, 75,9pp

 \item Sohn, S.T., Besla, G., van der Marel, R.P., et al. 2013, ApJ, 768, 139, 21pp

 \item Watkins, L.L., Evans, N.W., An, J.H. 2010, MNRAS, 406, 264-278

 \item Wilkinson M.I., Evans, N.W. 1999, MNRAS, 310, 645-662

 \item Xue, X.X., Rix, H.-W., Zhao, G., et al. 2008, ApJ, 684, 1143-1158

 \item The Hipparcos and Tycho Catalogues. 1997, ESA SP--1200

\end{itemize}
\end{document}